\documentclass[twocolumn,preprintnumbers,amsmath,amssymb]{revtex4}

\usepackage{graphicx}
\usepackage{dcolumn}
\usepackage{bm}
\usepackage{multirow}

\begin{document}
\title{Experimental implementation of heat-bath algorithmic cooling using solid-state nuclear magnetic resonance}

\author{Jonathan Baugh}
\email{baugh@iqc.ca}
\author{Osama Moussa}
\author{Colm A. Ryan}
\author{Ashwin Nayak}
\altaffiliation[Also at ]{Perimeter Institute for Theoretical
Physics, Waterloo, ON}
\author{Raymond Laflamme}
\email{rlaflamme@iqc.ca} \homepage{http://www.iqc.ca}
\altaffiliation[Also at ]{Perimeter Institute for Theoretical Physics, Waterloo, ON}
\affiliation{Institute for Quantum Computing, University of
Waterloo, Waterloo, Ontario N2L 3G1}

\date{Nov. 24, 2005}
\begin{abstract}
The counter-intuitive properties of quantum mechanics have the
potential to revolutionize information processing by enabling
efficient algorithms with no known classical counterparts
\cite{NC00a,EHC04a}. Harnessing this power requires developing a
set of building blocks \cite{DiV00a}, one of which is a method to
initialize the set of quantum bits (qubits) to a known state.
Additionally, fresh ancillary qubits must be available during the
course of computation to achieve fault tolerance \cite{KLZ98a,
Kit97a, AB97a, Pre98a}.  In any physical system used to implement
quantum computation, one must therefore be able to selectively and
dynamically remove entropy from the part of the system that is to
be mapped to qubits.  One such method is an "open-system" cooling
protocol in which a subset of qubits can be brought into contact
with an external large heat-capacity system. Theoretical efforts
\cite{SV90a, BMR+02a, FLMR04a} have led to an
implementation-independent cooling procedure, namely heat-bath
algorithmic cooling (HBAC). These efforts have culminated with the
proposal of an optimal algorithm, the partner-pairing algorithm
(PPA), which was used to compute the physical limits of HBAC
\cite{SMW05a}. We report here the first experimental realization
of multi-step cooling of a quantum system via HBAC. The experiment
was carried out using nuclear magnetic resonance (NMR) of a
solid-state ensemble three-qubit system.  It demonstrates the
repeated repolarization of a particular qubit to an effective
spin-bath temperature and alternating logical operations within
the three-qubit subspace to ultimately cool a second qubit below
this temperature. Demonstration of the control necessary for these
operations is an important milestone in the control of solid-state
NMR qubits and toward fault-tolerant quantum computing.
\end{abstract}

\maketitle

\section{Introduction}
\indent NMR-based ensemble quantum information processing (QIP)
devices have provided excellent testbeds for controlling
non-trivial numbers of qubits \cite{KLMT00a, GC97a, CPH98a,
CLK+00a}. A solid-state NMR QIP architecture builds on this
success by incorporating the essential features of the
liquid-state devices while offering the potential to reach unit
polarization and thus control more qubits \cite{CLK+00a,LOGM03a}.
In this architecture, the abundant nuclear spins with polarization
P form a large heat-capacity spin-bath that can be either coupled
to, or decoupled from, a dilute, embedded ensemble of
spin-labelled isotopomers that comprise the qubit register.  Bulk
spin-cooling procedures such as dynamic nuclear polarization are
well known and capable of reaching polarizations near unity
\cite{CLK+00a, AG82a}.  This architecture is one realization
within a large class of possible solid-state QIP systems in which
coherently controlled qubits can be brought into contact with an
external system that behaves as a heat bath.  The principles and
methods applied in solid-state NMR QIP will therefore apply to
many other systems.  An additional motivation is development of
control techniques that future quantum devices will utilize.  For
this experiment, we develop a novel technique to implement the
controlled qubit-bath interaction, and also report the first
application of strongly-modulating pulses \cite{FPB+02a} to
solid-state NMR for high-fidelity, coherent qubit control.
\section{Three-qubit malonic acid system}
\begin{figure}
\includegraphics[height=9.5cm]{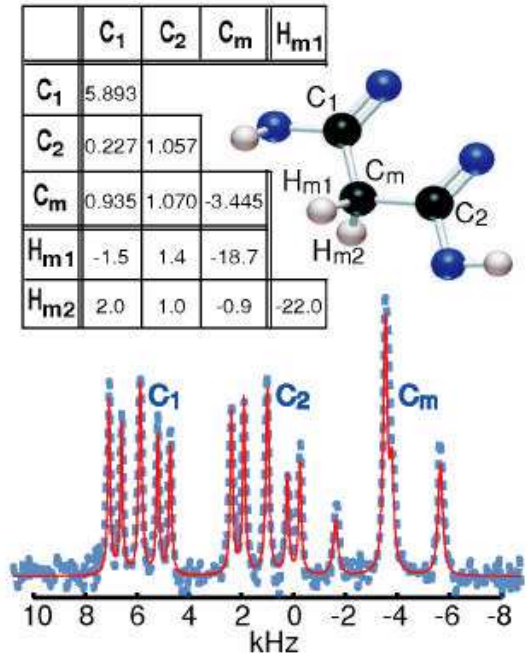}
\caption{Characteristics of the dilute $3$-$^{13}$C malonic acid
spin system. (below) $^{1}$H-decoupled, $^{13}$C spectrum near the
[010] orientation with respect to the static magnetic field. The
blue-dashed line is the experimental NMR absorption spectrum, and
the solid-red line is a fit. Multiplet assignments are indicated
by the labels $C_1$, $C_2$ and $C_m$. The central peaks in each
multiplet correspond to natural abundance $^{13}$C in the sample,
which are inconsequential for QIP purposes. The peak height
differences in the $3$-$^{13}$C molecule peaks indicate the strong
coupling regime, i.e. the $^{13}$C-$^{13}$C intramolecular dipolar
couplings are significant compared to the relative chemical
shifts. (above) Table showing the $^{13}$C rotating-frame
Hamiltonian parameters (chemical shifts along diagonal; dipolar
coupling strengths off-diagonal; all values in kHz) obtained from
the spectral fit. It also includes calculated dipolar couplings
involving the methylene protons based on the atomic coordinates
\cite{JRS94a} and the crystal orientation obtained from the
spectral fit. \label{fig:fig1}}
\end{figure}

\indent The three-qubit quantum information processor used here is
formed by the three spin-$1/2$ $^{13}$C nuclei of isotopically
labelled malonic acid molecules, occupying a dilute fraction of
lattice sites in an otherwise unlabeled single-crystal of malonic
acid (unlabeled, with the exception of naturally occurring
$^{13}$C isotopes at the rate of $1.1\%$). The concentration of
labelled molecules was $3.2\%$. Malonic acid also contains
abundant spin-$1/2$ $^{1}$H nuclei, which comprise the heat-bath.
Figure~\ref{fig:fig1} shows the $^{1}$H-decoupled, $^{13}$C-NMR
spectrum for the crystal (and crystal orientation) used in this
work. The spectrum shows the NMR absorption peaks of both the
qubit spins (quartets) and natural abundance $^{13}$C spins
(singlets), the latter being inconsequential for QIP purposes. The
table in Fig.~\ref{fig:fig1} lists the parameters of the ensemble
qubit Hamiltonian obtained from fitting the spectrum, and also
includes couplings involving the methylene protons calculated for
this crystal orientation from the known crystal structure
\cite{JRS94a}. Experiments were performed at room temperature at a
static magnetic field strength of $7.1$ T, where the thermal
$^{1}$H polarization is $P_{H}\simeq
2.4\times 10^{-5}$. \\
\section{Refresh operation}
\indent In this orientation, the methylene carbon $C_m$ has a
dipolar coupling of $19$ kHz to $H_{m1}$ of the methylene $^{1}$H
pair, whereas no other $^{13}$C-$^{1}$H dipolar coupling in the
system is larger than $2$ kHz. Therefore, a spin-exchange
Hamiltonian of the form\\
\begin{equation}
\mathcal{H}_{ex}=\sum_{j\in C,k\in
H}\frac{D_{jk}}{3}\frac{\sigma^{j}_{z}\sigma^{k}_{z}+\sigma^{j}_{y}\sigma^{k}_{y}+\sigma^{j}_{x}\sigma^{k}_{x}}{2}
\end{equation}
that couples the two nuclear species will generate dynamics
dominated by the large $C_m$-$H_{m1}$ coupling at short times (the
$D_{jk}$ are $^{13}$C-$^{1}$H dipolar couplings, indices run over
$^{13}$C,$^{1}$H nuclei, respectively, and
$\sigma^{\alpha}_{\beta}$ is the $\beta$-axis Pauli operator for
spin $\alpha$). Starting from the natural coupling Hamiltonian,\\
\begin{equation}
\mathcal{H}_{nat}=\sum_{j\in C,k\in
H}D_{jk}\sigma^{j}_{z}\sigma^{k}_{z}/2,
\end{equation}
we applied a multiple-pulse 'time-suspension' sequence
\cite{CMG90a} synchronously to both $^{13}$C and $^{1}$H spins to
create the effective spin-exchange Hamiltonian (in the toggling
frame), to lowest order in the Magnus expansion of the average
Hamiltonian \cite{Hae76a}. Application of the sequence for the
$C_m$-$H_{m1}$ exchange period $\tau=\frac{3}{4\times
19\text{kHz}}\simeq 40 \mu s$ results in an approximate swap gate
(state exchange) between the $C_m$ and $H_{m1}$ spins. With an
initial bulk $^{1}$H polarization $P_H$, this procedure yields a
selective dynamic transfer of polarization $P'=\eta P_H$ to $C_m$,
where $0\leq\eta\leq 1$ and ideally $|\eta|=1$ . We define the
effective spin-bath temperature to be that which corresponds to
the experimentally obtained $P'$ under this procedure, and refer
to this transfer as a refresh operation. We obtained $P'\simeq
0.83 P_H$ experimentally, and found that repeated refresh
operations showed no loss in efficiency given at least a $6$ ms
delay for $^{1}$H-$^{1}$H equilibration. However, we observed a
decay of $P_H$ as a function of the number of repetitions, due to
accumulated control errors, which lead to an identical loss in the
refresh polarization. \\
\section{Implementation of partner-pairing algorithm}
\begin{figure}
\includegraphics[height=8cm]{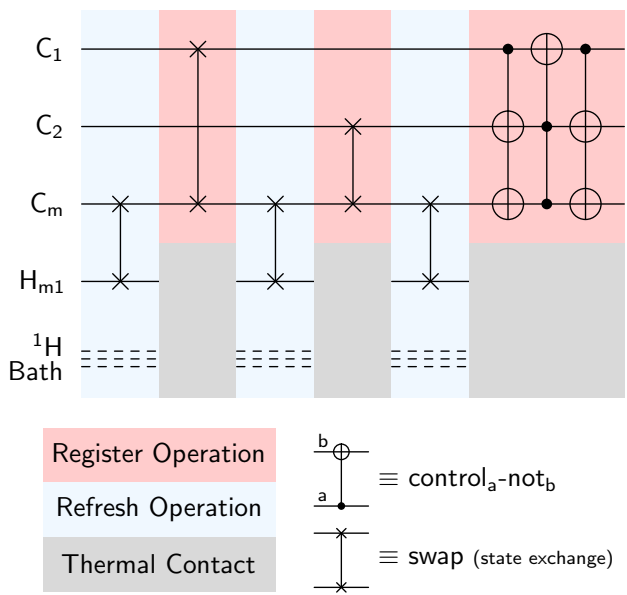}
\caption{Schematic quantum circuit diagram of the implemented
protocol.  Time flows from left to right. The three-bit
compression ($3$BC) gate is shown here decomposed as control-not
gates and a control-control-not (Toffoli) gate.  The gate sequence
corresponds to the first six steps of the partner-pairing
algorithm \cite{SMW05a} on three qubits.  The input state is a
collective polarization of the bulk $^{1}$H spins.  The refresh
operation is approximately $40\mu$s in duration, whereas the
register operations are between $0.7$ and $1.3$ ms in duration.
Thermal contact takes place during $^{1}$H spin-locking pulses
that begin just prior to the register operations, and extend an
additional $12$ ms after each operation.  $H_{m1}$ can be thought
of as an additional 'special purpose' qubit in this experiment;
despite non-selective $^{1}$H control (due to bulk hydrogenation),
the refresh and thermal contact operations could be performed
using collective $^{1}$H control. Thus, $H_{m1}$ serves as a
fast-relaxing 'qubit' and the bulk $^{1}$H-bath as a large
heat-capacity thermal bath. \label{fig:fig2}}
\end{figure}

\indent The experiment consists of the first six operations of the
partner-pairing algorithm (PPA) on three qubits: three refresh
operations, and three permutation gates that operate on the qubit
register.  This is described in the quantum circuit diagram of
Fig.~\ref{fig:fig2}.  During the register operations, the $^{1}$H
polarization is first rotated into the transverse plane, and then
'spin-locked' by a strong, phase-matched RF field that both
preserves the bulk $^{1}$H polarization and decouples the
$^{1}$H-$^{13}$C dipolar interactions.  Since $^{1}$H-$^{1}$H
dipolar interactions are merely scaled by a factor  $-1/2$ under
spin-locking, $H_{m1}$ is allowed to equilibrate with the bulk
$^{1}$H nuclei via spin diffusion. This occurs on a timescale
longer than the transverse dephasing time ($T_2(H_m)\sim 100
\mu$s), but much shorter than the spin-lattice relaxation time
($T^{H}_1\sim 50$s) of $H_{m1}$. Hence, $H_{m1}$ plays the role of
the fast-relaxing qubit described in the protocol of Schulman et
al. \cite{SMW05a}.  The first two register operations are swap
gates; the third is a three-bit compression ($3$BC) gate
\cite{SV90a,BMR+02a,FLMR04a} that boosts the polarization of the
first qubit, $C_1$, at the expense of the polarizations of the
other two qubits. Ideally, the protocol builds a uniform
polarization on all three qubits corresponding to the bath
polarization (first five steps), then selectively transfers as
much entropy as possible from the first qubit to the other two
(last step).  The last step ($3$BC) leads to a polarization boost
by a factor of $3/2$ on the first qubit. Subsequently, the heated
qubits can be re-cooled to the spin-bath temperature, and the
compression step repeated, iteratively, until the asymptotic value
of the first-bit polarization is reached. This limiting
polarization depends only on the number of qubits and the bath
polarization \cite{SMW05a}, and is ideally $P(C_1)=2P'$ for three
qubits (for $n$ qubits it is $2^{n-2}P'$ in the regime
$P'<<2^{-n}$, and $1.0$ in the regime $P'>>2^{-n}$
(refs.~\cite{SMW05a, Mou05a}). The first six steps carried out
here should yield a polarization of $1.5 P'$ on $C_1$, assuming
ideal operations. \\
\indent The control operations performed herein are quantum
control operations: state-independent unitary rotations in the
Hilbert space. However, it should be noted that the HBAC gates are
all permutations that map computational basis states to other
computational basis states. Therefore, gate fidelities were
measured with respect to correlation with these known states,
rather than the manifold of generic quantum states. We took
advantage of this property to further optimize the control
parameters of the $^{13}$C gates (register operations) for the
state-specific transformations of the protocol. These operations
were carried out using numerically optimized control sequences
referred to as strongly-modulating pulses \cite{FPB+02a}.  Such
pulses drive the system strongly at all times, such that the
average RF amplitude is comparable to, or greater than, the
magnitude of the internal Hamiltonian.  This allows
inhomogeneities in the ensemble qubit Hamiltonian to be
efficiently refocused, so that ensemble coherence is better maintained throughout the gate operations.\\
\indent In this set of experiments, the $^{13}$C qubit spins are
initialized to infinite temperature (a preceding broadband
$^{13}$C $\pi/2$ excitation pulse is followed by a dephasing
period in which $^{1}$H dipolar fields effectively dephase the
$^{13}$C polarization). Following the fifth step, polarizations
(in units of $P'$) of $0.88$, $0.83$ and $0.76$ ($\pm 0.03$) are
built up on $C_1$, $C_2$ and $C_m$, respectively. The final $3$BC
operation yields $P(C_1)/P'=1.22\pm 0.03$, a boost of $48\%$
compared to the average polarization ($0.82$) following step five.
Despite control imperfections that effectively heat the qubits at
each step, we are able to cool the $C_1$ qubit ensemble well below
the effective $^{1}$H spin-bath temperature.\\
\section{Results}
\begin{figure}
\includegraphics[height=10cm]{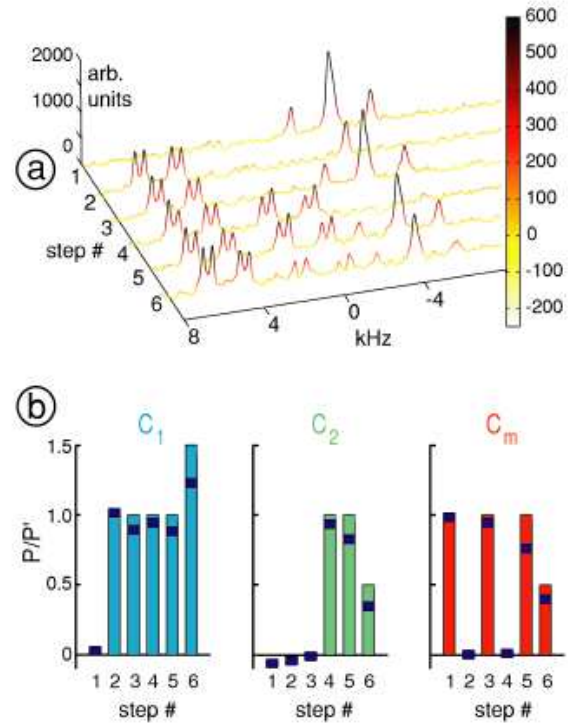}
\caption{Experimental results in terms of $^{13}$C spectra and
their integrated intensities. (a) Readout spectra obtained
following each of the six steps in the protocol. The integrated
peak intensities for each multiplet correspond to the ensemble
spin polarizations.  The natural abundance $C_m$ signal that
appears at each refresh step (adding to the intensity of the
central peaks) should be ignored; we are only interested in the
part of the signal arising from the $3$-$^{13}$C qubit molecules,
which can be seen clearly in the $C_1$ and $C_2$ spectral regions.
(b) Bars indicate ideal qubit polarizations at each step;
experimental values obtained from integration of the above spectra
are shown as shaded bands, whose thickness indicates experimental
uncertainty.\label{fig:fig3}}
\end{figure}

\indent The results are summarized in Fig.~\ref{fig:fig3}; in (a)
are shown the spectral intensities corresponding to $^{13}$C spin
polarizations following each of the six steps, and in (b) the
integrated intensities are graphed in comparison with the ideal
values.  We note that the overall fidelity of the experiment,
$F=1.22/1.50=0.81$, implies an error per step of $3.7\%$.  This
error rate is only about a factor of two larger than the average
error per two-qubit gate obtained in a benchmark liquid-state NMR
QIP experiment \cite{KLMT00a}. Furthermore, the state-correlation
fidelity of the $3$BC gate over the polarizations on all three
qubits is $0.96\pm 0.03$. From Fig.~\ref{fig:fig3}(b), it can be
seen that the fidelity of the refresh operation drops off roughly
quadratically in the number of steps; this is consistent with the
loss of bulk $^{1}$H polarization due to pulse imperfections both
in the multiple-pulse refresh operations and in the spin-locking
sequence.  Since the broadband pulses have been optimized for
flip-angle in these sequences, we suspect that the remaining
errors are mainly due to switching transients that occur in the
tuned RF circuitry of the NMR probehead, and to a lesser extent
off-resonance and finite pulse-width effects that modify the
average Hamiltonian \cite{CMG90a}. Similar effects lead to
imperfect fidelity of the $^{13}$C control.  With suitable
improvements to the resonant circuit response and by incorporating
numerical optimization of the multiple-pulse refresh operations,
we expect that several iterations of the protocol could be carried
out and that the limiting polarization of $2P'$ could be
approached in this system. The same methodologies should also be
applicable in larger qubit systems with similar architecture.  For
a $6$-qubit system using the PPA, a bath polarization $P>0.2$
would be sufficient, in principle, to reach a pure state on one
qubit \cite{Mou05a}. Such bulk nuclear polarizations are well
within reach via well-known dynamic nuclear polarization
techniques \cite{AG82a}; for example, unpaired electron spins at
defects ($g=2$) in a field of $3.4$ T and at temperature $4.2$ K are polarized to $0.5$.\\
\indent This work demonstrates that solid-state NMR QIP devices
could be used to implement active error correction.  Given a bath
polarization near unity, the refresh operation implemented here
would constitute the dynamic resetting of a chosen qubit. This
would allow a new NMR-based testbed for the ideas of quantum error
correction and for controlled open-system quantum dynamics in the
regime of high state purity and up to $\sim 20$ qubits.\\

\section{\label{sec:level1}Methods}
NMR experiments were carried out at room temperature on a Bruker
Avance solid-state spectrometer operating at a field of $7.1$ T,
and home-built dual channel RF probe. The sample coil had an inner
diameter of $3$ mm, and the employed $\pi/2$ broadband pulse
lengths were $1.25\mu$s and $0.75\mu$s for $^{13}$C and $^{1}$H,
respectively. The sample was a $4.5\times 1\times 1$ mm$^3$ single
crystal of malonic acid grown from aqueous solution with a $3.2\%$
molecular fraction of $3-$$^{13}$C labelled molecules. Spectra
were obtained by signal averaging for $80$ scans. The proton
spin-lattice relaxation time was $T^{H}_1=50$s, so the delay
between scans was set to $6T^{H}_1=300$s. Design of the
strongly-modulating $^{13}$C pulses followed very closely the
methodology described in ref.~\cite{FPB+02a}, and penalty
functions were adjusted to favor average RF amplitudes comparable
to or greater than the magnitude of the $^{13}$C rotating-frame
Hamiltonian. These pulses were optimized and simulated over a
$5$-point distribution of RF amplitude corresponding to the
measured distribution over the spin ensemble ($\sigma=6.2\%$ in RF
amplitude). The 'time-suspension' sequence applied synchronously
to $^{13}$C and $^{1}$H was a $12$-pulse subsequence of the Cory
$48$-pulse sequence \cite{CMG90a}. The delays between pulses were
adjusted so that the total length of the sequence was $40\mu$s.
$^{1}$H spin-locking/decoupling was carried out at RF amplitude of
$250$ kHz. The spectra in Fig.~\ref{fig:fig3} were obtained by
applying a $\pi/2$ broadband pulse to read out the spin
polarizations. The readout pulses were preceded by a $1.5$ ms
delay in which $^{1}$H decoupling was on but any off-diagonal
terms in the $^{13}$C spin density matrix would significantly
dephase ($T^{\ast}_2\simeq 2$ms). The absolute value of the
refresh polarization $P'$ was determined by comparing the initial
refresh polarization on $C_m$ and the thermal equilibrium $^{13}$C
polarization $P_C$ measured in a separate experiment.  These yield
the ratio of $P'$ to $P_C$, and $P'$ to $P_H$ using the fact that
$P_H=3.98 P_C$.

\small{We gratefully acknowledge: D. G. Cory, T. F. Havel, and C.
Ramanathan for discussions and use of NMR simulation code; W. P.
Power, M. Ditty and N. J. Taylor for facility use and experimental
assistance; ARDA and NSERC for support.  O. M. acknowledges the
Ontario Ministry of Training, Colleges and Universities for
support.}\\
\small{Correspondence and requests for material should be sent to
J. B. (e-mail: baugh@iqc.ca).}


\begin{thebibliography}{22}
\expandafter\ifx\csname
natexlab\endcsname\relax\def\natexlab#1{#1}\fi
\expandafter\ifx\csname bibnamefont\endcsname\relax
  \def\bibnamefont#1{#1}\fi
\expandafter\ifx\csname bibfnamefont\endcsname\relax
  \def\bibfnamefont#1{#1}\fi
\expandafter\ifx\csname citenamefont\endcsname\relax
  \def\citenamefont#1{#1}\fi
\expandafter\ifx\csname url\endcsname\relax
  \def\url#1{\texttt{#1}}\fi
\expandafter\ifx\csname
urlprefix\endcsname\relax\def\urlprefix{URL }\fi
\providecommand{\bibinfo}[2]{#2}
\providecommand{\eprint}[2][]{\url{#2}}

\bibitem[{\citenamefont{Nielsen and Chuang}(2000)}]{NC00a}
\bibinfo{author}{\bibfnamefont{M.~A.} \bibnamefont{Nielsen}} \bibnamefont{and}
  \bibinfo{author}{\bibfnamefont{I.~L.} \bibnamefont{Chuang}},
  \emph{\bibinfo{title}{Quantum Computation and Quantum Information}}
  (\bibinfo{publisher}{Cambridge University Press},
  \bibinfo{address}{Cambridge, UK}, \bibinfo{year}{2000}).

\bibitem[{EHC(2004)}]{EHC04a}
\bibinfo{journal}{Quant. Inform. Process.} \textbf{\bibinfo{volume}{3}},
  \bibinfo{pages}{1} (\bibinfo{year}{2004}), \bibinfo{note}{special issue}.

\bibitem[{\citenamefont{DiVincenzo}(2000)}]{DiV00a}
\bibinfo{author}{\bibfnamefont{D.~P.} \bibnamefont{DiVincenzo}},
  \bibinfo{journal}{Fort. der Phys.} \textbf{\bibinfo{volume}{48}},
  \bibinfo{pages}{771} (\bibinfo{year}{2000}), \eprint{quant-ph/0002077}.

\bibitem[{\citenamefont{Knill et~al.}(1998)\citenamefont{Knill, Laflamme, and
  Zurek}}]{KLZ98a}
\bibinfo{author}{\bibfnamefont{E.}~\bibnamefont{Knill}},
  \bibinfo{author}{\bibfnamefont{R.}~\bibnamefont{Laflamme}}, \bibnamefont{and}
  \bibinfo{author}{\bibfnamefont{W.~H.} \bibnamefont{Zurek}},
  \bibinfo{journal}{Science} \textbf{\bibinfo{volume}{279}},
  \bibinfo{pages}{342} (\bibinfo{year}{1998}).

\bibitem[{\citenamefont{Kitaev}(1997)}]{Kit97a}
\bibinfo{author}{\bibfnamefont{A.~Y.} \bibnamefont{Kitaev}}, in
  \emph{\bibinfo{booktitle}{Quantum Communication, Computing and Measurement}}
  (\bibinfo{publisher}{Plenum}, \bibinfo{address}{New York, USA},
  \bibinfo{year}{1997}), pp. \bibinfo{pages}{181--188}.

\bibitem[{\citenamefont{Aharonov and {Ben-Or}}(1997)}]{AB97a}
\bibinfo{author}{\bibfnamefont{D.}~\bibnamefont{Aharonov}} \bibnamefont{and}
  \bibinfo{author}{\bibfnamefont{M.}~\bibnamefont{{Ben-Or}}}, in
  \emph{\bibinfo{booktitle}{Proc. 29th. Ann. ACM Symp. on Theory of Computing}}
  (\bibinfo{year}{1997}), \bibinfo{note}{longer version quant-ph/9906129},
  \eprint{quant-ph/9611025}.

\bibitem[{\citenamefont{Preskill}(1998)}]{Pre98a}
\bibinfo{author}{\bibfnamefont{J.}~\bibnamefont{Preskill}},
  \bibinfo{journal}{Proc. R. Soc. Lond. A} \textbf{\bibinfo{volume}{454}},
  \bibinfo{pages}{385} (\bibinfo{year}{1998}), \eprint{quant-ph/9705031}.

\bibitem[{\citenamefont{Schulman and Vazirani}(1999)}]{SV90a}
\bibinfo{author}{\bibfnamefont{L.}~\bibnamefont{Schulman}} \bibnamefont{and}
  \bibinfo{author}{\bibfnamefont{U.}~\bibnamefont{Vazirani}},
  \bibinfo{journal}{Proc. of the 31th Annual ACM Symposium on Theory of
  Computing} pp. \bibinfo{pages}{322--329} (\bibinfo{year}{1999}).

\bibitem[{\citenamefont{Boykin et~al.}(2002)\citenamefont{Boykin, Mor,
  Roychowdhury, Vatan, and Vrijen}}]{BMR+02a}
\bibinfo{author}{\bibfnamefont{P.~O.} \bibnamefont{Boykin}},
  \bibinfo{author}{\bibfnamefont{T.}~\bibnamefont{Mor}},
  \bibinfo{author}{\bibfnamefont{V.}~\bibnamefont{Roychowdhury}},
  \bibinfo{author}{\bibfnamefont{F.}~\bibnamefont{Vatan}}, \bibnamefont{and}
  \bibinfo{author}{\bibfnamefont{R.}~\bibnamefont{Vrijen}},
  \bibinfo{journal}{Proc. Natl. Acad. Sci. USA} \textbf{\bibinfo{volume}{99}},
  \bibinfo{pages}{3388} (\bibinfo{year}{2002}), \eprint{quant-ph/0106093}.

\bibitem[{\citenamefont{Fernandez et~al.}(2004)\citenamefont{Fernandez, Lloyd,
  Mor, and Roychowdhury}}]{FLMR04a}
\bibinfo{author}{\bibfnamefont{J.~M.} \bibnamefont{Fernandez}},
  \bibinfo{author}{\bibfnamefont{S.}~\bibnamefont{Lloyd}},
  \bibinfo{author}{\bibfnamefont{T.}~\bibnamefont{Mor}}, \bibnamefont{and}
  \bibinfo{author}{\bibfnamefont{V.}~\bibnamefont{Roychowdhury}},
  \bibinfo{journal}{Int. J. Quant. Inform.} \textbf{\bibinfo{volume}{2}},
  \bibinfo{pages}{461} (\bibinfo{year}{2004}).

\bibitem[{\citenamefont{Schulman et~al.}(2005)\citenamefont{Schulman, Mor, and
  Weinstein}}]{SMW05a}
\bibinfo{author}{\bibfnamefont{L.}~\bibnamefont{Schulman}},
  \bibinfo{author}{\bibfnamefont{T.}~\bibnamefont{Mor}}, \bibnamefont{and}
  \bibinfo{author}{\bibfnamefont{Y.}~\bibnamefont{Weinstein}},
  \bibinfo{journal}{Phys. Rev. Lett.} \textbf{\bibinfo{volume}{94}},
  \bibinfo{pages}{120501} (\bibinfo{year}{2005}).

\bibitem[{\citenamefont{Knill et~al.}(2000)\citenamefont{Knill, Laflamme,
  Martinez, and Tseng}}]{KLMT00a}
\bibinfo{author}{\bibfnamefont{E.}~\bibnamefont{Knill}},
  \bibinfo{author}{\bibfnamefont{R.}~\bibnamefont{Laflamme}},
  \bibinfo{author}{\bibfnamefont{R.}~\bibnamefont{Martinez}}, \bibnamefont{and}
  \bibinfo{author}{\bibfnamefont{C.-H.} \bibnamefont{Tseng}},
  \bibinfo{journal}{Nature} \textbf{\bibinfo{volume}{404}},
  \bibinfo{pages}{368} (\bibinfo{year}{2000}), \eprint{quant-ph/9908051},
  \urlprefix\url{http://www.ariv.org/abs/quant-ph/9908051}.

\bibitem[{\citenamefont{Gershenfeld and Chuang}(1997)}]{GC97a}
\bibinfo{author}{\bibfnamefont{N.}~\bibnamefont{Gershenfeld}} \bibnamefont{and}
  \bibinfo{author}{\bibfnamefont{I.~L.} \bibnamefont{Chuang}},
  \textbf{\bibinfo{volume}{275}}, \bibinfo{pages}{350} (\bibinfo{year}{1997}).

\bibitem[{\citenamefont{Cory et~al.}(1998)\citenamefont{Cory, Price, and
  Havel}}]{CPH98a}
\bibinfo{author}{\bibfnamefont{D.~G.} \bibnamefont{Cory}},
  \bibinfo{author}{\bibfnamefont{M.~D.} \bibnamefont{Price}}, \bibnamefont{and}
  \bibinfo{author}{\bibfnamefont{T.~F.} \bibnamefont{Havel}},
  \bibinfo{journal}{Physica D} \textbf{\bibinfo{volume}{120}},
  \bibinfo{pages}{82} (\bibinfo{year}{1998}), \eprint{quant-ph/9709001}.

\bibitem[{\citenamefont{Cory et~al.}(2000)\citenamefont{Cory, Laflamme, Knill,
  Viola, Havel, N.Boulant, Boutis, Fortunato, Lloyd, Martinez
  et~al.}}]{CLK+00a}
\bibinfo{author}{\bibfnamefont{D.~G.} \bibnamefont{Cory}},
  \bibinfo{author}{\bibfnamefont{R.}~\bibnamefont{Laflamme}},
  \bibinfo{author}{\bibfnamefont{E.}~\bibnamefont{Knill}},
  \bibinfo{author}{\bibfnamefont{L.}~\bibnamefont{Viola}},
  \bibinfo{author}{\bibfnamefont{T.~F.} \bibnamefont{Havel}},
  \bibinfo{author}{\bibnamefont{N.Boulant}},
  \bibinfo{author}{\bibfnamefont{G.}~\bibnamefont{Boutis}},
  \bibinfo{author}{\bibfnamefont{E.}~\bibnamefont{Fortunato}},
  \bibinfo{author}{\bibfnamefont{S.}~\bibnamefont{Lloyd}},
  \bibinfo{author}{\bibfnamefont{R.}~\bibnamefont{Martinez}},
  \bibnamefont{et~al.}, \bibinfo{journal}{Fort. der Phys. special issue,
  Experimental Proposals for Quantum Computation} \textbf{\bibinfo{volume}{48}}
  (\bibinfo{year}{2000}), \eprint{quant-ph/0004104}.

\bibitem[{\citenamefont{Leskowitz et~al.}(2003)\citenamefont{Leskowitz, Olsen,
  N.Ghaderi, and Mueller}}]{LOGM03a}
\bibinfo{author}{\bibfnamefont{G.~M.} \bibnamefont{Leskowitz}},
  \bibinfo{author}{\bibfnamefont{R.~A.} \bibnamefont{Olsen}},
  \bibinfo{author}{\bibnamefont{N.Ghaderi}}, \bibnamefont{and}
  \bibinfo{author}{\bibfnamefont{L.~J.} \bibnamefont{Mueller}},
  \bibinfo{journal}{J. Chem. Phys.} \textbf{\bibinfo{volume}{119}},
  \bibinfo{pages}{1643} (\bibinfo{year}{2003}).

\bibitem[{\citenamefont{Abragam and Goldman}(1982)}]{AG82a}
\bibinfo{author}{\bibfnamefont{A.}~\bibnamefont{Abragam}} \bibnamefont{and}
  \bibinfo{author}{\bibfnamefont{M.}~\bibnamefont{Goldman}},
  \emph{\bibinfo{title}{Nuclear Magnetism: Order and Disorder}}
  (\bibinfo{publisher}{Oxford University Press}, \bibinfo{address}{Oxford,
  England}, \bibinfo{year}{1982}).

\bibitem[{\citenamefont{Fortunato et~al.}(2002)\citenamefont{Fortunato, Pravia,
  Boulant, Teklemariam, Havel, and Cory}}]{FPB+02a}
\bibinfo{author}{\bibfnamefont{E.~M.} \bibnamefont{Fortunato}},
  \bibinfo{author}{\bibfnamefont{M.~A.} \bibnamefont{Pravia}},
  \bibinfo{author}{\bibfnamefont{N.}~\bibnamefont{Boulant}},
  \bibinfo{author}{\bibfnamefont{G.}~\bibnamefont{Teklemariam}},
  \bibinfo{author}{\bibfnamefont{T.~F.} \bibnamefont{Havel}}, \bibnamefont{and}
  \bibinfo{author}{\bibfnamefont{D.~G.} \bibnamefont{Cory}},
  \bibinfo{journal}{J. Chem. Phys.} \textbf{\bibinfo{volume}{116}},
  \bibinfo{pages}{7599} (\bibinfo{year}{2002}).

\bibitem[{\citenamefont{Jagannathan et~al.}(1994)\citenamefont{Jagannathan,
  Rajan, and Subramanian}}]{JRS94a}
\bibinfo{author}{\bibfnamefont{N.~R.} \bibnamefont{Jagannathan}},
  \bibinfo{author}{\bibfnamefont{S.~S.} \bibnamefont{Rajan}}, \bibnamefont{and}
  \bibinfo{author}{\bibfnamefont{E.}~\bibnamefont{Subramanian}},
  \bibinfo{journal}{J. Chem. Cryst.} \textbf{\bibinfo{volume}{24}},
  \bibinfo{pages}{75} (\bibinfo{year}{1994}).

\bibitem[{\citenamefont{Cory et~al.}(1990)\citenamefont{Cory, Miller, and
  Garroway}}]{CMG90a}
\bibinfo{author}{\bibfnamefont{D.~G.} \bibnamefont{Cory}},
  \bibinfo{author}{\bibfnamefont{J.~B.} \bibnamefont{Miller}},
  \bibnamefont{and} \bibinfo{author}{\bibfnamefont{A.~N.}
  \bibnamefont{Garroway}}, \bibinfo{journal}{J. Mag. Res.}
  \textbf{\bibinfo{volume}{90}}, \bibinfo{pages}{205} (\bibinfo{year}{1990}).

\bibitem[{\citenamefont{Haeberlen}(1976)}]{Hae76a}
\bibinfo{author}{\bibfnamefont{U.}~\bibnamefont{Haeberlen}},
  \emph{\bibinfo{title}{High Resolution {NMR} in Solids: Selective Averaging}}
  (\bibinfo{publisher}{Academic Press}, \bibinfo{address}{New York, USA},
  \bibinfo{year}{1976}).

\bibitem[{\citenamefont{Moussa}(2005)}]{Mou05a}
\bibinfo{author}{\bibfnamefont{O.}~\bibnamefont{Moussa}},
  \emph{\bibinfo{title}{On heat-bath algorithmic cooling and its implementation
  in solid-state {NMR}}} (\bibinfo{year}{2005}), \bibinfo{note}{{MSc.} thesis},
  \eprint{http://www.iqc.ca/~omoussa/work/thesis}.

\end{thebibliography}
\end{document}